\documentclass[10pt,conference]{IEEEtran}
\IEEEoverridecommandlockouts
\usepackage{hyperref}
\usepackage[square,sort&compress,numbers]{natbib}

\usepackage{amsmath,amsfonts}
\usepackage{algorithmic}
\usepackage{graphicx}
\usepackage[table]{xcolor}
\usepackage{enumitem}
\usepackage{textcomp}
\usepackage[T1]{fontenc}
\usepackage{listings}
\usepackage{float}
\usepackage[font=small,skip=0pt]{caption}
\usepackage[linesnumbered,ruled]{algorithm2e}
\usepackage{threeparttable}
\usepackage{subcaption}
\usepackage{examplep}
\usepackage{booktabs}
\usepackage{pifont}
\usepackage{setspace}
\usepackage{cprotect}
\usepackage{todonotes}
\usepackage{framed}
\usepackage{tcolorbox}
\usepackage{multirow}
\usepackage{adjustbox}
\usepackage{fancyvrb}
\usepackage{wrapfig}
\usepackage{verbatimbox}
\usepackage{siunitx}
\usepackage{stmaryrd}
\def\BibTeX{{\rm B\kern-.05em{\sc i\kern-.025em b}\kern-.08em
    T\kern-.1667em\lower.7ex\hbox{E}\kern-.125emX}}
    
    \makeatletter
\let\old@lstKV@SwitchCases\lstKV@SwitchCases
\def\lstKV@SwitchCases#1#2#3{}
\makeatother
\usepackage{lstlinebgrd}
\makeatletter
\let\lstKV@SwitchCases\old@lstKV@SwitchCases

\lst@Key{numbers}{none}{%
    \def\lst@PlaceNumber{\lst@linebgrd}%
    \lstKV@SwitchCases{#1}%
    {none:\\%
     left:\def\lst@PlaceNumber{\llap{\normalfont
                \lst@numberstyle{\thelstnumber}\kern\lst@numbersep}\lst@linebgrd}\\%
     right:\def\lst@PlaceNumber{\rlap{\normalfont
                \kern\linewidth \kern\lst@numbersep
                \lst@numberstyle{\thelstnumber}}\lst@linebgrd}%
    }{\PackageError{Listings}{Numbers #1 unknown}\@ehc}}
\makeatother
    
\begin{document}

\newcommand{\TODO}[1]{\textcolor{red}{TODO: #1}}
\newcommand{\NOTE}[1]{\textcolor{blue}{NOTE: #1}}
\newcommand{\CITE}[0]{\textcolor{orange}{[CITE]}}
\newcommand\APPROACH{{\sf Comprex}}
\newcommand\BFAPPROACH{{\sf \textbf{Comprex}}}
\newcommand\ITAPPROACH{{\sf \textit{Comprex}}}
\newcommand\BFITAPPROACH{{\sf \textbf{\textit{Comprex}}}}
\newcommand{\cmark}{\ding{51}}%
\newcommand{\xmark}{\ding{55}}%
\newcommand\colorgr{green!40}
\newcommand\colorbl{blue!25}
\newcommand{\gr}{\cellcolor{\colorgr}}
\newcommand{\bl}{\cellcolor{\colorbl}}    
\newcommand\constr[1]{\llbracket #1 \rrbracket}

\SetKwProg{Fn}{Function}{}{end}	
\SetAlFnt{\small}
\SetAlCapFnt{\small}
\SetAlCapNameFnt{\small}
\SetKwBlock{When}{when}{end}
\SetKwBlock{Until}{until}{end}

\newcommand\mycommfont[1]{\textit{#1}}
\SetCommentSty{mycommfont}

\appto\TPTnoteSettings{\footnotesize}    

\newsavebox{\savelisting}
\newenvironment{listing}
{\begin{lrbox}{\savelisting}
\begin{minipage}{3.5in}
\begin{flushleft}}
{\end{flushleft}
\end{minipage}
\end{lrbox}
\begin{center}
\resizebox{\columnwidth}{!}{\setlength\fboxsep{6pt}\fbox{\usebox{\savelisting}}}
\end{center}}

\lstdefinestyle{mystyle}{%
  frame            = tblr,    
  tabsize          = 1,     
  numbers          = left,  
  numbersep = 4pt,  
  framesep         = 3pt,   
  framerule        = 0pt, 
  commentstyle     = \color{darkgray},      
  keywordstyle     = \color{blue},       
  showstringspaces = false,              
  escapeinside={<@}{@>}
}

\title{White-Box Analysis over Machine Learning: Modeling Performance of Configurable Systems}

\makeatletter
\newcommand{\linebreakand}{%
  \end{@IEEEauthorhalign}
  \hfill\mbox{}\par
  \mbox{}\hfill\begin{@IEEEauthorhalign}
}
\makeatother

\author{\IEEEauthorblockN{Miguel Velez}
\IEEEauthorblockA{Carnegie Mellon University}
\and
\IEEEauthorblockN{Pooyan Jamshidi}
\IEEEauthorblockA{University of South Carolina}
\and
\IEEEauthorblockN{Norbert Siegmund}
\IEEEauthorblockA{Leipzig University}
\linebreakand
\IEEEauthorblockN{Sven Apel}
\IEEEauthorblockA{Saarland Informatics Campus}
\IEEEauthorblockA{Saarland University}
\and
\IEEEauthorblockN{Christian Kästner}
\IEEEauthorblockA{Carnegie Mellon University}
}

\IEEEaftertitletext{\vspace{-2\baselineskip}}

\maketitle
\pagestyle{plain}

\begin{abstract}
Performance-influence models can help stakeholders understand how and where configuration options and their interactions influence the performance of a system.
With this understanding, stakeholders can 
debug performance behavior and make deliberate 
configuration decisions. 
Current black-box techniques to build such models combine various sampling and learning strategies, resulting in tradeoffs between measurement effort, accuracy, and interpretability.  
We present \BFAPPROACH, a white-box approach to build performance-influence models for configurable systems, combining insights of local measurements,
dynamic taint analysis to track options in the implementation, compositionality, and compression of the configuration space, without relying on machine learning to extrapolate incomplete samples.
Our evaluation on 4 widely-used, open-source projects demonstrates that \BFAPPROACH\ builds similarly accurate performance-influence models 
to the most accurate and expensive black-box approach, but 
at a reduced cost and with additional benefits from interpretable and local models.
\looseness=-1
\end{abstract}


\section{Introduction}
\label{intro-sec}


Configuring software systems is often challenging.
In practice, many users execute systems with inefficient configurations in terms of performance and, often directly correlated, energy consumption~\cite{HY:ESEM16, JSSSL:PLDI12, HDGZX:ICSE12, NJT:MSR13}.
While users can adjust configuration options to tradeoff between performance and the system's functionality, this configuration task can be overwhelming; many systems, such as databases, Web servers, and video encoders, have numerous configuration options that may interact, possibly producing unexpected and undesired behavior.
For this reason, understanding how options and their interactions affect the system's performance and making suitable configuration decisions can be difficult~\cite{XJFZPT:ESECFSE15, ABKS:FOSPL13}; users often have, at most, vague intuitions~\cite{XZHZSYZP:SOSP13, HXC:VAMOS12,SGAK:ESECFSE15}.
We aim at efficiently building performance-influence models, without machine learning, that characterize how options influence the system's performance.
We intent our models to be easy to inspect and interpret by developers and users to ease performance debugging and to make informed configuration decisions.
\looseness=-1

Good performance-influence models help users understand \emph{how}, \emph{where}, and \emph{why} options and their interactions influence the performance of the system in a specific way~\cite{KSKGA:SOSYM18,SGAK:ESECFSE15}.
In the situations where these models are useful, simple manual experiments are often not practical due to a system's exponentially large configuration space. 
For example, a developer considering whether to enable encryption would likely expect that encryption would slow down the system.
However, the developer may need to perform \emph{numerous} experiments to quantify the effect of the option itself and of its interactions with other configuration decisions in a given environment and workload.
To understand why and where encryption causes the slow down, a developer would have to carefully profile or debug the system in different configurations.
\looseness=-1

\begin{figure}[t]
\begin{center}
    \includegraphics[width=\columnwidth]{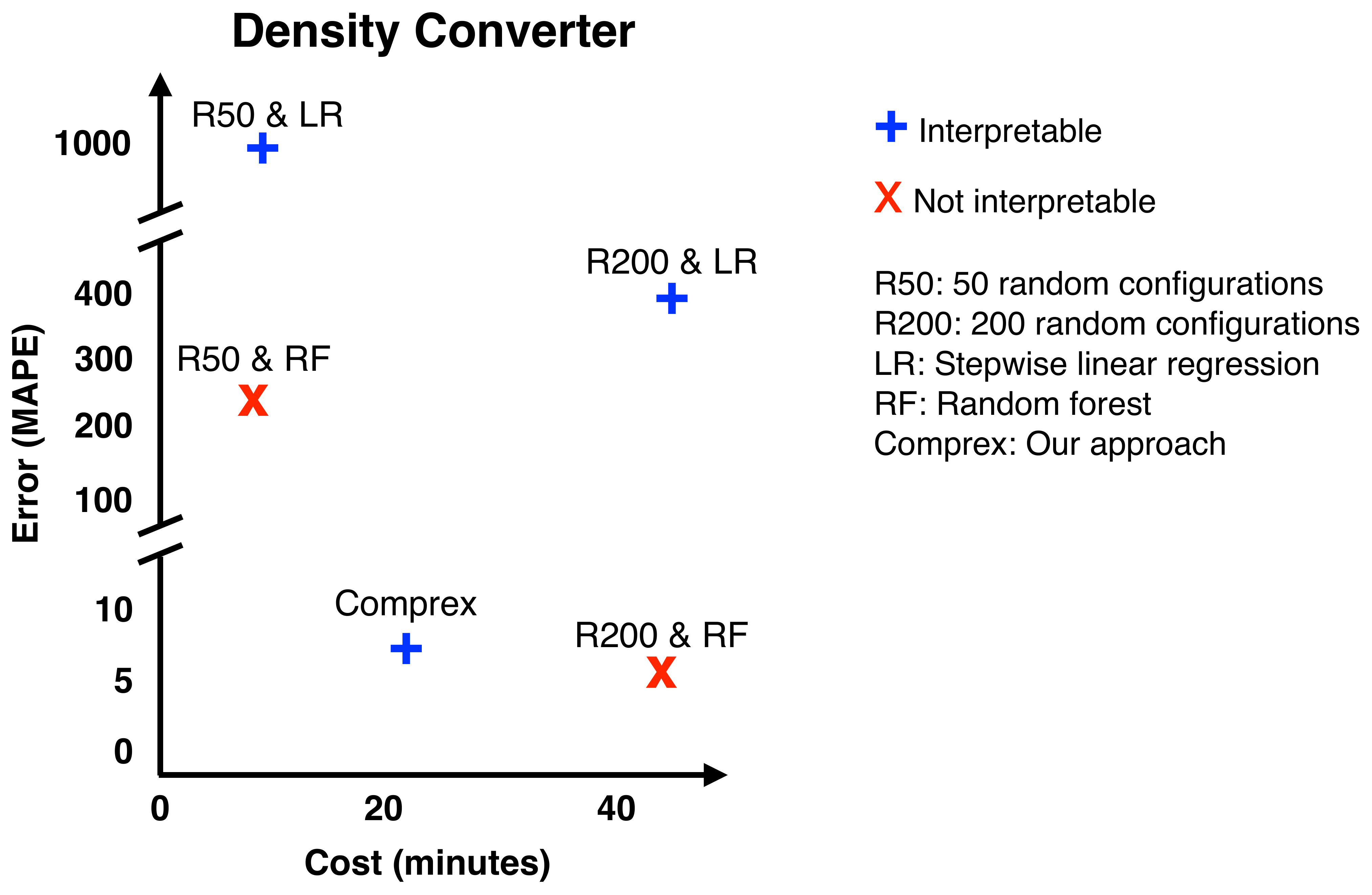}
\end{center}
\caption{
Our white-box approach, \APPROACH, builds similarly accurate (low error) performance-influence models to the most accurate and expensive sampling and machine learning approaches, but can often be built more efficiently (low cost) and are interpretable.
Additionally, our models are significantly more accurate than other sampling and machine learning approaches that produce interpretable linear models. 
\looseness=-1
}
\label{mape-cost-interpret-h2-fig}
\end{figure}

Most existing techniques build global performance-influence models and treat the system as a \emph{black box}, 
measuring the system's execution in 
an environment with a given workload for a subset of all configurations, and learning a model from these observations.
The sampling (i.e., selecting which configurations to measure) and learning techniques used~\cite{GCASW:ASE13, SGSAC:ASE15, SKKABRS:ICSE12, SRKKAS:SQJ12, SGAK:ESECFSE15,KGSGA:ICSE19, M:IML20, GSA:PPSCNSB19, KGSA:IEEESOFT20} result in tradeoffs among 
the cost to build the models and the accuracy and interpretability of the models~\cite{GSA:PPSCNSB19, KGSA:IEEESOFT20, KSKGA:SOSYM18}.
For example, larger samples are more expensive, but usually lead to more accurate models; random forests, with large enough samples, tend to learn more accurate models than those built with linear regression, but the models are harder to interpret when users want to understand performance or debug their systems~\cite{M:IML20, GSA:PPSCNSB19, KGSA:IEEESOFT20} (see Fig.~\ref{mape-cost-interpret-h2-fig}).
\looseness=-1

In this paper, we introduce \APPROACH, a \emph{white-box} approach to build \emph{accurate}, \emph{interpretable}, and \emph{local} performance-influence models at low cost.
Our approach departs drastically from traditional black-box approaches: It analyzes and instruments the source code to accurately capture configuration-specific performance behavior, without relying on machine learning to extrapolate incomplete samples.
We reduce measurement cost by \emph{simultaneously analyzing and measuring} multiple code regions of the system, building a local linear performance-influence model per region with a few configurations (an insight we call \emph{compression}).
Subsequently, we \emph{compose} the local models into a global model for the entire system.
We use an \emph{iterative dynamic taint analysis} to identify \emph{where} and \emph{how} load-time configuration options influence control-flow statements in the system, through \emph{control-flow and data-flow dependencies}. 
\looseness=-1

Our empirical evaluation on $4$ medium- to large-scale, widely-used, open-source projects demonstrates that \APPROACH\ builds similarly accurate performance-influence models to those learned with the most accurate and expensive black-box approach, but at a 
reduced cost.
Furthermore, \APPROACH\ generates \emph{interpretable} and \emph{local} models, which 
quantify the influence of individual options and interactions on performance, and can even map the influence to code regions.
\looseness=-1

In summary, we make the following contributions:
\begin{itemize}[noitemsep,topsep=0pt]
\item An approach that combines \emph{compression} and \emph{composition} to accurately infer the influence of options on the performance of numerous independent regions of a system with a few configurations, composing the influence in a full performance model.
\looseness=-1
\item An iterative dynamic taint analysis to identify how configurations influence the performance of independent regions and a conjecture to reduce the cost of running the analysis to build accurate performance-influence models.
\looseness=-1
\item An empirical evaluation on $4$ medium- to large-scale, 
open-source systems demonstrating that our white-box approach builds interpretable, but similarly accurate models with less cost compared to the state of the art.
\looseness=-1
\item A replication package with subject systems, experimental setup, 
and data of several months of measurements~\cite{VJSAK:ICSE21SM}.
\looseness=-1
\end{itemize}

\section{Performance Models for Config. Systems}
\label{state-of-the-art-sec}

There is substantial literature on modeling the performance of software systems~\cite[e.g.,][]{VN:PMAS15,GSA:PPSCNSB19,KGSA:IEEESOFT20,KSKGA:SOSYM18}.
Performance-influence models solve a specific problem: Explaining how options and their interactions influence a system's performance for a given workload and environment, designed to help users understand performance and make deliberate configuration decisions.
\looseness=-1

Traditional performance models are typically used by designers and developers to model and analyze the performance of a system's architecture (e.g., using Queuing networks, Petri Nets, and Stochastic Process Algebras) and workload in the \emph{design stage} of a project~\cite{H:PMDCSQTA13, SCBSCB:QEST06, K:PMEDCSUQPN06}.
At this stage, design decisions are usually modeled as configuration options \cite{EEM:TSE13, BKR:JSS09}.
By contrast, performance-influence models describe the performance behavior of a \emph{system implementation} in a given workload and environment, and are typically \emph{learned} by observing the behavior of the \emph{system implementation}.
\looseness=-1

In the context of configurable systems, performance-influence models \emph{predict} a system's performance in terms of configuration options and their interactions~\cite{SGAK:ESECFSE15, GCASW:ASE13,JSVKPA:ASE17,JVKS:FSE18, VPGFC:ICPE17, VJSSAK:ASEJ20,JVKSK:SEAMS17,KSKGA:SOSYM18,JVKS:FSE18,HZ:ICSME19,HZ:ICSE19}.
Some models are explicitly designed to \emph{explain} how specific options and their interactions influence a system's performance~\cite{SKKABRS:ICSE12,SRA:GPCE13,GCASW:ASE13,SGAK:ESECFSE15,VJSSAK:ASEJ20,KSKGA:SOSYM18}.
For example, the sparse linear model $8+15\texttt{A}+10\texttt{C}+3\texttt{AB}+30\texttt{AC}$ captures the execution time of the system in Fig.~\ref{running-example-fig}, which 
predicts the performance of arbitrary configurations, and explains \emph{how} the options \texttt{A}, \texttt{B}, and \texttt{C} and their interactions influence the system's performance.
The model can be inspected by developers, for example, to determine whether the large performance impact when combining \texttt{A} and \texttt{C} is an unexpected performance bug.
\looseness=-1

\begin{figure}[t]
\lstset{style=mystyle}
\begin{lstlisting}[language=Java,
morekeywords={def},
						 tabsize = 2,
						 numbers = left, 
						 % escapechar = ~, 
						 basicstyle = \ttfamily\scriptsize, 
						 % linewidth = .6\linewidth,
						 mathescape = true,
						 firstnumber = auto,
						 name = wordpress,
						 xleftmargin = 2em,
						 framexleftmargin=1.5em,
						 linebackgroundcolor={%
                                                        \ifnum\value{lstnumber}>0
                                                                \ifnum\value{lstnumber}<16
                                                                        \color{yellow!20}
                                                                \fi
                                                        \fi
                                                        \ifnum\value{lstnumber}>15
                                                                \ifnum\value{lstnumber}<19
                                                                        \color{blue!20}
                                                                \fi
                                                        \fi
                                                        \ifnum\value{lstnumber}>18
                                                                \ifnum\value{lstnumber}<22
                                                                        \color{red!20}
                                                                \fi
                                                        \fi}                   
						 ]
def main(List workload) 
    a = getOpt("A"); b = getOpt("B"); $\label{r:loadoptions1}$
    c = getOpt("C"); d = getOpt("D"); $\label{r:loadoptions2}$
    ...         // execution: 1s
    int i = 0;               <@T$_{main}$ = 3 - 1A@>
    if(a)       // variable depends on option A $\label{r:ifa}$
        ...     // execution: 1s      
        foo(b); // variable depends on option B
        i = 20;                  
    else
        ...     // execution: 2s
        i = 5;
    while(i > 0) $\label{r:whilei}$
        bar(c); // variable depends on option C
        i--;
def foo(boolean x)           <@T$_{foo}$ = 1A + 3AB@>$\label{r:iffoox}$
    if(x) ...   // execution: 4s
    else ...    // execution: 1s 
def bar(boolean x)           <@T$_{bar}$ = 5 + 15A + 10C + 30AC@>$\label{r:ifbarx}$
    if(x) ...   // execution: 3s
    else ...    // execution: 1s    
\end{lstlisting}
\caption{Example system with three colored regions (methods) influenced by configuration options. Local performance-influence models for a specific workload are shown per region.\looseness=-1}
\label{running-example-fig}
\end{figure}

\paragraph*{Building performance-influence models} 
Performance-influence models are typically built by 
measuring a system's execution time with the target workload in the target environment under different configurations~\cite{SGAK:ESECFSE15}.
Almost all existing approaches are \emph{black-box} in nature: They do not analyze the system's implementation and measure the end-to-end execution time of the 
system.
\looseness=-1

The simplest approach is to observe the execution of all configurations in a \emph{brute-force} approach.
The approach obviously does not scale,
as the number of configurations grows exponentially with the number of options.
\looseness=-1

In practice, most approaches measure executions only for a \emph{sampled} subset of all configurations and  extrapolate performance behavior for the rest of the configuration space using machine learning~\cite{GSA:PPSCNSB19}, which we collectively refer to as \emph{sampling and learning} approaches.
Specific approaches differ in how they sample, learn, and represent models: Common sampling techniques include uniform random, feature-wise, and pair-wise sampling~\cite{MKRGA:ICSE16}, design of experiments~\cite{M:DAE06}, and combinatorial sampling~\cite{AKTLS:GPCE16,NL:CSUR11,HBG:SRE11,HMGB:IST16,HNADPB:ESE18}.
Common learning techniques include linear regression~\cite{SKKABRS:ICSE12, SRKKAS:SQJ12, SGAK:ESECFSE15, KGSGA:ICSE19}, regression trees~\cite{GCASW:ASE13, SGSAC:ASE15, GYSASVCWY:ESE17, GSA:PPSCNSB19},  Fourier Learning~\cite{HZ:ICSME19},
and neural networks~\cite{HZ:ICSE19}.
Different sampling and learning techniques yield different tradeoffs between  measurement  effort, prediction  accuracy, and interpretability of the learned models~\cite{GSA:PPSCNSB19, KGSA:IEEESOFT20, KSKGA:SOSYM18}.
\looseness=-1

\paragraph*{Goals of performance-influence modeling}
\label{goals-subsec}
Performance-influence models can be used for different tasks in different scenarios, which benefit from different model characteristics.
\looseness=-1

In the simplest case, a user wants to \emph{optimize} a system's performance by selecting the fastest configuration for their workload and in the target environment.
Performance-influence models have been used for optimization~\cite{OBMS:ESECFSE17, ZLGBMLSY:SOCC17, GCASW:ASE13, NMSA:ESECFSE17}, though metaheuristic search (e.g., hill climbing) is often more effective at pure optimization problems~\cite{HHL:LION11, OBMS:ESECFSE17, ORGC:SPLC14, JC:MASCOTS16}, as they do not need to understand the entire configuration space.
\looseness=-1

In other scenarios, users want to predict the performance of individual configurations.
Scenarios include \emph{automatic reconfiguration} and \emph{runtime adaptation}, where there is no human in the loop and online search is impractical.
For example, when dynamically deciding during a robot's mission which options to change to react to
low battery levels~\cite{WLHLSK:ASPLOS18, ZLGBMLSY:SOCC17, JVKSK:SEAMS17, JVKS:FSE18}.
In these scenarios, the model's prediction accuracy is important across the entire configuration space, but understanding the structure of the model is not important.
In this context, deep regression trees~\cite{GCASW:ASE13, SGSAC:ASE15, GYSASVCWY:ESE17}, Fourier Learning~\cite{HZ:ICSME19}, and neural networks~\cite{HZ:ICSE19} are commonly used, which build accurate models, but are not easy to interpret by humans~\cite{SGAK:ESECFSE15, KGSA:IEEESOFT20, KSKGA:SOSYM18, GSA:PPSCNSB19,M:IML20}.
\looseness=-1

When performance-influence models are used by users to make deliberate configuration decisions~\cite{SGAK:ESECFSE15, WLHLSK:ASPLOS18, XZHZSYZP:SOSP13, KGSA:IEEESOFT20, KSKGA:SOSYM18, GSA:PPSCNSB19} (e.g., whether to accept the performance overhead of encryption), interpretability regarding \emph{how} options and interactions influence performance becomes paramount.
In these settings, researchers usually suggest sparse linear models, such as $8+15\texttt{A}+10\texttt{C}+3\texttt{AB}+30\texttt{AC}$ above, typically learned with stepwise linear regression or similar variations~\cite{SKKABRS:ICSE12, SRKKAS:SQJ12, SGAK:ESECFSE15, KGSGA:ICSE19}.
In such models, users can directly recognize the influence of an option or an interaction.
\looseness=-1

In addition to end users who configure a system, developers who maintain the system can also benefit from performance-influence models to understand and debug the performance behavior of their systems~\cite{KSKGA:SOSYM18, VJSSAK:ASEJ20, KIJRJ:CADET20}.
For example, when presenting performance-influence models to developers in high-performance computing, 
\citet{KSKGA:SOSYM18}
reported that a developer  \emph{``was surprised to see that [an option] had only a small influence on system performance''}, indicating a potential bug.
In such setting, understanding \emph{how} individual options and interactions influence performance is again paramount, favoring interpretable models.
Ideally, models would also indicate \emph{where} the influence occurs in the implementation. 
For example, in prior work~\cite{VJSSAK:ASEJ20}, we shared experience of how mapping performance influence of options to specific code regions was useful in identifying several inconsistencies between documentation and implementation.
\looseness=-1

In this paper, we build performance models that can be used in all of these scenarios. The models are \textit{accurate} for optimization and prediction, they are interpretable explaining \textit{how} options and interactions influence performance, and they map performance influence to code regions explaining \emph{where} performance influence manifests.
Our white-box approach, \APPROACH, builds these models by analyzing and observing system internals, without using machine learning.
\looseness=-1


\section{Comprex}
\label{approach-sec}

We introduce \BFAPPROACH, a \emph{white-box} approach to efficiently and accurately generate performance-influence models, without using traditional sampling and learning techniques.
\looseness=-1

Similar to black-box approaches, we build performance-influence models by observing the execution of a system in different configurations, but we guide the exploration with a white-box analysis of the internals of the system.
For a given input, a system with a set of Boolean options $O$ can exhibit up to $2^{|O|}$ distinct execution paths, one per configuration.\footnote{
For simplicity, we describe \APPROACH\ in terms of Boolean options, but other option types can be encoded and discretized.
The distinction between inputs and options is subjective and domain specific. 
We consider options as 
special 
inputs with a small finite domain (e.g., Boolean) that a user might explore to change functionality or influence quality attributes.
We consider fixed values for other inputs. 
Note that a user might fix some configuration options
and consider alternative values for inputs
(e.g., use an option for different workloads). 
However, we 
explore the performance influence of options with finite domains, assuming all other inputs are fixed at specific values.
This setting results in a finite, but typically very large configuration space.
\looseness=-1
}
If we measure the execution time of each distinct path, we can build a performance-influence model that accurately maps performance differences to options and their interactions, without any approximation through machine learning.
The resulting models can be expressed as familiar linear 
models.
\looseness=-1

\begin{figure}[t]
\begin{center}
\includegraphics[width=\columnwidth]{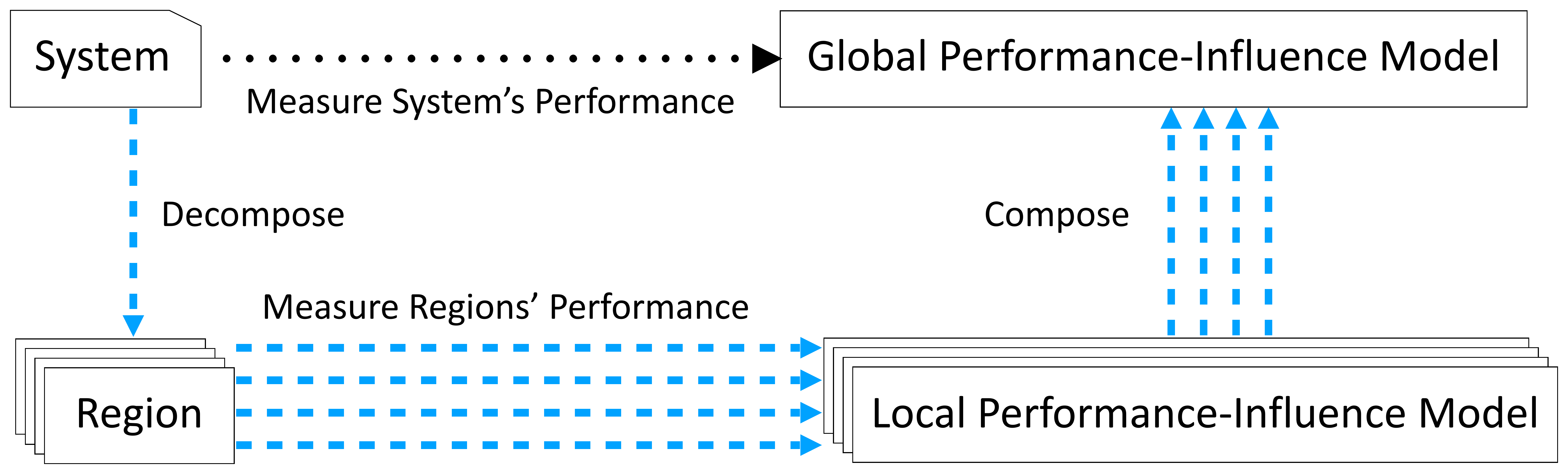}
\end{center}
\caption{Building performance-influence models is compositional: Instead of building a single model for the entire system (dotted black arrow), we can simultaneously build a local model per region and compose those models (dashed blue arrows).
\looseness=-1
}
\label{compresion-fig}
\end{figure}

Our approach to efficiently build performance-influence models relies on two insights: ($1$)~Performance-influence models can be built \emph{compositionally}, by composing models built independently for smaller regions of the code (cf. Fig.~\ref{compresion-fig}). 
($2$)~Multiple performance-influence models for smaller regions can be built simultaneously by observing a system's executions often with only a few configurations, which we call \emph{compression}.
\looseness=-1

First, building performance-influence models is compositional: We can measure the performance of smaller regions in the system (e.g., considering each method as a region) and build a performance-influence model per region separately, which describes each region's performance behavior in terms of options.
Subsequently, we can compose the local models to describe the performance of the entire system, computed as the sum of the individual influences in each model (e.g., composing $5 + 4\texttt{A}$ and $1 - 1\texttt{A} + 2\texttt{B}$ to $6 + 3\texttt{A} + 2\texttt{B}$).
\looseness=-1

Compositionality helps reduce the cost of our approach, as many smaller regions of a system are often influenced only by a subset of all options, confirmed by prior empirical research~\cite{MWKTS:ASE16, NKCFP:FSE16}.
Hence, the number of distinct paths to observe in a region is usually much smaller than the number of distinct paths in the entire system.
If we have an analysis that can identify the subset of options that directly and indirectly influence the smaller region (see Sec.~\ref{idta-sec}), we can build a local performance-influence model by observing all distinct paths in a region often with only a few configurations.
\looseness=-1

Second, compression makes our approach scale without relying on machine learning approximations: When executing a single configuration, we can \emph{simultaneously} observe the execution of multiple regions.
In addition, if the regions are influenced by different options, a common case confirmed by prior
research~\cite{MWKTS:ASE16, NKCFP:FSE16}, then we can separately measure, \emph{in one configuration}, the influence of different options in different regions in the system, instead of exploring all combinations of all options.
For example, three independent regions \texttt{if(a)\{\} if(b)\{\} if(c)\{\}} influenced by options \texttt{A}, \texttt{B}, and \texttt{C}, respectively, each have two distinct paths.
Instead of exploring all $8$ combinations of the three options, we can explore all distinct paths in each region with only $2$ configurations, as long as each option is enabled in one configuration and disabled in the other configuration.
\looseness=-1

\begin{figure}[t]
\begin{center}
\includegraphics[width=\columnwidth]{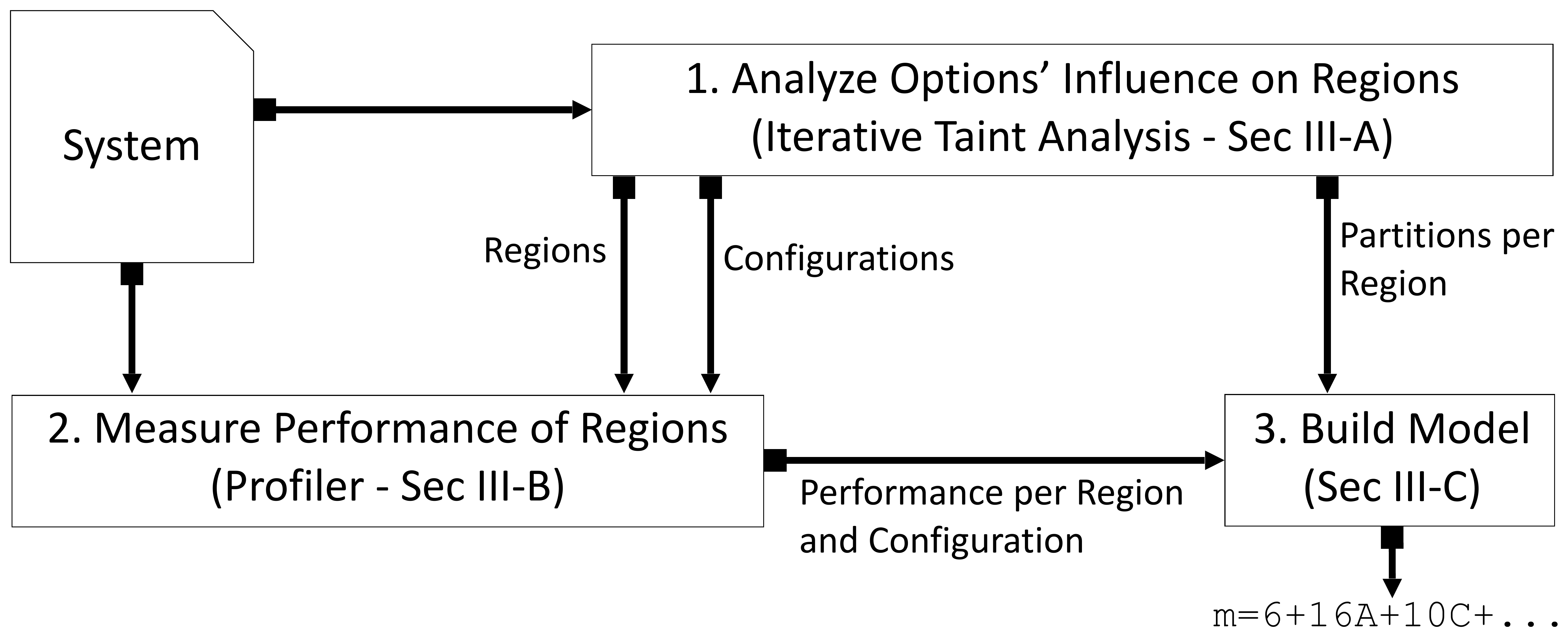}
\end{center}
\caption{Overview of \APPROACH's three components.}
\label{approach-fig}
\end{figure}

Our approach combines compositionality and compression to build accurate performance-influence models, without traditional sampling or machine-learning techniques. 
The resulting models 
can be presented in an 
interpretable format
and even be mapped to individual code regions.
Key to our approach is the property that 
not all options interact in the same region, instead influencing different parts of the system independently; a pattern observed empirically in configurable systems~\cite{MWKTS:ASE16, NKCFP:FSE16}.
\looseness=-1

To operationalize compression and compositionality for building accurate and interpretable performance-influence models with low cost, we need three technical components, as illustrated in Fig.~\ref{approach-fig}: 
First, we identify which regions are influenced by which options to select configurations to explore all paths per region and to map measured execution times to options and their interactions~(Sec.~\ref{idta-sec}).
Second, we execute the system to measure the performance of all regions~(Sec.~\ref{profile-sec}).
Third, we build local performance-influence models per region and compose them into one global model for the entire system~(Sec.~\ref{build-sec}).
\looseness=-1

\subsection{Analyzing Options' Influence on Regions}
\label{idta-sec}

As a first step, we identify which options (directly or indirectly) influence control-flow decisions in which code regions.
We use this information to select configurations to explore all paths per region and map measured performance differences to options and their interactions (Sec.~\ref{profile-sec}).\footnote{
We focus on different execution paths caused by configuration changes, 
fixing all other inputs.
We focus on configuration changes in control-flow statements, as a system's execution time changes 
in those statements, depending on which branch is executed and how many times it is executed, confirmed by empirical research~\cite{HY:ESEM16, NCRL:ICSE15, VJSSAK:ASEJ20, SRA:GPCE13, JSSSL:PLDI12}.
Execution differences caused by nondeterminism are orthogonal and must be handled in conventional ways (e.g., averaging multiple observations or controlling the environment).
\looseness=-1
}
To this end, we track \emph{information flow} from configuration options (sources) to \emph{control-flow decisions} (sinks) in each region.
If a configuration option flows, directly or indirectly (including implicit flows), into a control-flow decision in a region, this implies that selecting or deselecting the option may lead to different execution paths within the region.
Thus, we should observe at least one execution with a configuration in which the option is selected and another execution in which the option is not selected.
\looseness=-1

More specifically, we conservatively partition the configuration space per region into subspaces, such that every configuration in each subspace takes the same path through the control-flow decisions within a region, and that all distinct paths are explored when taking one configuration from each subspace.
Formally, a set of Boolean options $O$ forms a configuration space $C=\mathcal{P}(O)$ where each configuration $c\in C$ is represented by the set of selected options.
A \emph{partition} of the configuration space ($p\subseteq\mathcal{P}(C)$) is a grouping of configurations into nonempty subsets, which we call \emph{subspaces}, such that each configuration is part of exactly one subspace.
For notational convenience, we describe subspaces using propositional formulas over options.
For example, $\constr{\texttt{A}\wedge\neg\texttt{B}}$ describes the subspace of all configurations in which option \texttt{A} is selected and option \texttt{B} is deselected.
\looseness=-1

To track information flow between options and control-flow decisions in regions, we employ a \emph{dynamic taint analysis} and \emph{iteratively} collect information through repeated executions of the system in different configurations, until we have explored all distinct paths in each region.
A taint analysis, typically used in the security domain~\cite{BK:SIGPLANNOT14, ARFBBKLOM:PLDI14}, tracks how values are affected by selected inputs (sources) and are used in specific locations (sinks). 
During each execution, we track how API calls load configuration options (Lines~\ref{r:loadoptions1}--\ref{r:loadoptions2} in Fig.~\ref{running-example-fig}) (sources) and propagate them along data-flow and control-flow dependencies, including implicit flows, to the decisions of control-flow statements (sinks).
We use a dynamic rather than a static analysis~\cite{LKB:TSE18, VJSSAK:ASEJ20}, since it scales to larger systems and avoids some overtainting problems by tracking actual executions~\cite{BK:SIGPLANNOT14}.
\looseness=-1

\begin{algorithm}[t]
\scriptsize
\caption{Iterative Dynamic Taint Analysis}
\label{idta-algo}
	\KwIn{configurable system $p$}
	\KwOut{partition for each region $R \rightarrow \mathcal{P}(\mathcal{P}(C))$}
	\SetKwFunction{FIDTA}{partion\_all\_regions($p$)}
		
	\Fn{\FIDTA} {
		\textit{partitions} := $R \rightarrow \{C\}$; \textit{executed\_configs} := $\emptyset$ 
		
			
		\Until(\texttt{explored\_all\_subspaces(}\textit{executed\_configs}, \textit{partitions}\texttt{)} \textbf{do}) {\label{alg:mainloop}
			$cc$ := \texttt{get\_next\_config(}\textit{executed\_configs}, \textit{partitions}\texttt{)}\label{alg:getnextconfig} 	
		
			\textit{executed\_configs} := $\textit{executed\_configs} \cup cc$
			
			\textbf{during} \texttt{execute\_taint\_analysis($p, cc$)} \When(reaching a control-flow decision with taints $t_d$ and $t_c$ in region $r$:) {				
				
				\textit{partitions}\texttt{[$r$]} := $ \textit{partitions}\texttt{[$r$]}\times \texttt{get\_part($t_d, t_c, cc$)}$\label{alg:updatepartition2}
			}
		}
	
		\Return{\textit{partitions}}
	}
	
	\vspace{2ex}

    \KwIn{executed configurations $\textit{ec} : \mathcal{P}(C), \textit{partitions} : R \rightarrow \mathcal{P}(\mathcal{P}(C))$}
	\KwOut{\texttt{true} or \texttt{false}}
	\SetKwFunction{FExploredAllSubspaces}{explored\_all\_subspaces($ec$, $partitions$)}
	
		\Fn{\FExploredAllSubspaces} {
		
		$\textit{all\_subspaces}$ := $\bigcup \texttt{image}(\textit{partitions})$ 
			
		\Return{$\forall s \in \textit{all\_subspaces}. \ \exists c \in ec. \ c \in s$}
	}

	\vspace{2ex}
	
	\KwIn{data-flow taints $t_d$, control-flow taints $t_c$, current 
	configuration $cc$}
	\KwOut{partition $p : \mathcal{P}(\mathcal{P}(C))$ for the current decision}
	\SetKwFunction{FGetPartition}{get\_part($t_d, t_c, cc$)}
	
		\Fn{\FGetPartition} {
		
		$s_{reach}$ := $\{c\in C\,|\, \forall o\in t_c.\ \ o\in c \Leftrightarrow o\in cc \}$\label{alg:getpartitionreach} 
		
		$p$ := $\bigl\{C \setminus s_{reach}\bigr\}$\label{alg:getpartitionrest} 		\tcp{subspace of config. that might not reach decision}
		
		\tcp{add one subspace for every combination of options in data-flow taints}
		\For{$a \in \mathcal{P}(t_d)$}{\label{alg:getpartitiondata1} 
            $s$ := $\bigl\{c \in C\, | \, 
		   \forall o \in t_d. \ \ o \in c \Leftrightarrow o\in a\  \bigr\}$ \label{alg:getpartitiondata2}
		   
		   $p$ := $p \cup \{ s\cap s_{reach} \}$\label{alg:getpartitionintersectreach} 
		}  

		\Return{$p \setminus \emptyset$}
	}
\end{algorithm}

\paragraph*{Incrementally partitioning the configuration space}
Alg.~\ref{idta-algo} describes how we partition the configuration space per region, based on incremental updates from our dynamic taint analysis.
Intuitively, we execute the system in a configuration and observe when data-flow and control-flow taints from options reach each control-flow decision in each region, and subsequently update each region's partition:
Whenever we reach a control-flow statement during execution, we identify, based on taints that reach the condition of the statement, the sets of configurations that would possibly make different decisions,
thus updating the partition that represents different paths for this region (Line~\ref{alg:updatepartition2}).
Since a dynamic taint analysis can only track information flow in the current execution, but not for alternative executions (i.e., for paths not taken), we repeat the process with new configurations, selected from the partitions identified in prior executions, updating partitions until we have explored one configuration from each subspace of each partition (main loop, Line~\ref{alg:mainloop}); that is, until we have observed each distinct path in each region at least once.
Note that some subspaces in the region might make the same control-flow decision as other subspaces, but we do not know which subspace will make which decision until we actually execute those configurations.
\looseness=-1

Updating partitions works as follows:
When we reach a control-flow statement in a region with data-flow taints $t_d$, 
this indicates that the options in $t_d$ affect the control-flow decision, \emph{but other options do not}.
Thus, we know that all configurations that share the same selection for all options in $t_d$ will result in the same control-flow decision, while configurations with different selections of these options may result in different decisions.
Since the taint analysis tells us only that the options in $t_d$ may somehow (directly or indirectly) affect the decision's condition, but not how, we will need to explore at least one configuration for every possible assignment to these options, even though multiple or even all may end up taking the same branch.
Therefore, we partition the configuration space \emph{at this decision} corresponding to all combinations of the options in $t_d$ (Lines~\ref{alg:getpartitiondata1}--\ref{alg:getpartitiondata2}).
For example, for a decision influenced by options \texttt{A} and \texttt{B}, we partition the configuration space into four subspaces: all configurations in which \texttt{A} and \texttt{B} are selected together, all configurations in which \texttt{A} is selected but not \texttt{B}, all configurations in which \texttt{B} is selected but not \texttt{A}, and all configurations in which neither \texttt{A} nor \texttt{B} are selected.
Finally, we update the \emph{region's} partition with the partition derived for the decision by computing their cross product ($\times$, Line~\ref{alg:updatepartition2}).
This operation reflects that, to explore all paths among multiple control-flow decisions in a region, including multiple executions of the same control-flow statement, we need to explore all combinations of the individual paths in the region.
\looseness=-1

Distinguishing data-flow taints from control-flow taints allows us 
to optimize the exploration of nested decisions (e.g., \texttt{if(a)\{ if(b) ... \}}).
Control-flow taints specify which options (directly or indirectly) influenced outer control-flow decisions, which indicates that different assignments to options in the control-flow taints \emph{may} lead to paths where the current decision is not reached in the first place.
Hence, we do not necessarily need to explore all interactions of options affecting outer and inner decisions.
Instead of exploring combinations for all options of data-flow and control-flow taints, we first split the configuration space into ($1$)~those configurations for which we know that they will reach the current decision, as they share the assignments of options in control-flow taints ($s_\textit{reach}$, Line~\ref{alg:getpartitionreach}), and ($2$)~the remaining configurations which may not reach the current decision ($C\setminus s_\textit{reach}$, Line~\ref{alg:getpartitionrest}).
Then, we only create subspaces for interactions of options in data-flow taints within $s_\textit{reach}$ (Lines~\ref{alg:getpartitiondata1}--\ref{alg:getpartitionintersectreach}) and consider the entire set of configurations outside $s_\textit{reach}$ as a single subspace (Line~\ref{alg:getpartitionrest}).
The iterative nature of our analysis ensures that at least one of the configurations outside $s_\textit{reach}$ will be explored, and, if the configuration also reaches the same decision, the region's partition will be further divided.
\looseness=-1

The iterative analysis executes the system in different configurations until one configuration from each subspace of each partition in each region has been explored.
That is, we start by executing any 
configuration (e.g., the default configuration),
which reveals the subspaces per regions that could make different decisions.
The algorithm then selects the next configuration to explore unseen subspaces in the regions (Line~\ref{alg:getnextconfig}), which may further update the regions' partitions.
To select the next configuration, we use a greedy algorithm to pick a configuration that explores the most unseen subspaces across all regions.\footnote{
To avoid enumerating an exponential number of configurations, we use a greedy algorithm that picks a random subspace and incrementally intersects it with other non-disjoint subspaces, which seems sufficiently effective in practice.
The problem can also be encoded as a MAXSAT problem, representing subspaces as propositional formulas, to find the configuration that satisfies the formula with the most subspaces.
\looseness=-1
}
\looseness=-1

\paragraph*{Example}
We exemplify running the iterative analysis in our running example from Fig.~\ref{running-example-fig}, considering each method as a region.
If we execute 
the configuration $\{$\texttt{A},\texttt{D}$\}$, in which options \texttt{A} and \texttt{D} are selected and the other options are deselected, the taint analysis will indicate that the value of the variable \texttt{a} is tainted by option \texttt{A} (from Line~\ref{r:loadoptions1}) when reaching the \texttt{if} statement in Line~\ref{r:ifa} (region \texttt{main}).
Thus, 
all configurations in which \texttt{A} is selected result in the same control-flow decision and all configurations in which \texttt{A} is not selected result potentially in the same or a different decision.
Hence, we derive the initial partition $\{\constr{\texttt{A}}, \constr{\neg\texttt{A}}\}$ for this method.
\looseness=-1

Continuing the execution,
we next reach the \texttt{if} statement in Line~\ref{r:iffoox} (region \texttt{foo}), where the value of the variable \texttt{x} is tainted with control-flow taint \texttt{A} (from Line~\ref{r:ifa}) and data-flow taint \texttt{B} (from variable \texttt{b}).
Thus, 
all configurations in which \texttt{A} and \texttt{B} are selected result in the same control-flow decision, all configurations in which \texttt{A} is selected and \texttt{B} is not selected may result in a different control-flow decision, and 
all configurations in which \texttt{A} is not selected may not reach this decision.
Hence, we
derive the partition  $\{\constr{\texttt{A}\wedge\neg\texttt{B}}, \constr{\neg \texttt{A}}, \constr{\texttt{A}\wedge\texttt{B}}\}$ for this region. 
Note how we explore this nested \texttt{if} statement with $3$ instead of $4$ subspaces by separately tracking 
data-flow and control-flow taints. 
\looseness=-1

Further in the execution, the decision in the \texttt{while} statement (Line~\ref{r:whilei}) depends on the tainted value of the variable \texttt{i} (implicit flow), in each loop iteration, resulting in $\{\constr{\texttt{A}}$, $\constr{\neg\texttt{A}}\}$, which is 
consistent with \texttt{main}'s existing partition.
Hence, the cross product does not change the partition.
Similarly, the decision in Line~\ref{r:ifbarx} (region \texttt{bar}) repeatedly depends on data-flow taint \texttt{C} and control-flow taint \texttt{A}, resulting in the partition $\{\constr{\texttt{A}\wedge\neg\texttt{C}}, \constr{\neg\texttt{A}}, \constr{\texttt{A}\wedge\texttt{C}}\}$.
\looseness=-1

After this first execution, 
we identified six distinct subspaces among the partitions of the three regions ($\constr{\texttt{A}},\constr{\neg\texttt{A}},\constr{\texttt{A}\wedge\texttt{B}}, \constr{\texttt{A}\wedge\neg\texttt{B}},\constr{\texttt{A}\wedge\texttt{C}}$, and $\constr{\texttt{A}\wedge\neg\texttt{C}}$), of which $\constr{\texttt{A}}, \constr{\texttt{A}\wedge\neg\texttt{B}}$, and $\constr{\texttt{A}\wedge\neg\texttt{C}}$ were explored with the initial configuration.
In the next iteration, we select a configuration, for example $\{$\texttt{A},\texttt{B},\texttt{C}$\}$, to explore unseen subspaces in the regions and update partitions.
In this case, however, nothing changes.
We continue executing new configurations to explore 
unseen subspaces, possibly updating the regions' partitions, until we have explored all subspaces in the regions.
After executing only $4$ out of $16$ configurations, for example $\{\texttt{A},\texttt{D}\}$, $\{\texttt{A},\texttt{B},\texttt{C}\}$, $\{\}$, and $\{\texttt{C}\}$, we have explored at least one configuration from each subspace of each partition in each region, and the iterative analysis terminates.
The subspaces derived for the three regions's partitions are
$\constr{\texttt{A}}$, $\constr{\neg\texttt{A}}$, $\constr{\texttt{A}\wedge\neg\texttt{B}}$, $\constr{\texttt{A}\wedge\texttt{B}}$, $\constr{\neg\texttt{A}\wedge\neg\texttt{C}}$, $\constr{\neg\texttt{A}\wedge\texttt{C}}$, $\constr{\texttt{A}\wedge\neg\texttt{C}}$, $\constr{\texttt{A}\wedge\texttt{C}}$.
\looseness=-1

\paragraph*{Discussion}
Note how the iterative analysis explores regions independently and does not explore paths for options that do not influence that region (e.g., we do not explore the interaction of \texttt{B} and \texttt{C} and never explore configurations specifically for \texttt{D}).
Also, note how the 
taint analysis tracks both direct and indirect dependencies interprocedurally.
\looseness=-1

The iterative analysis is guaranteed to terminate, as it explores new configurations during each iteration.
In the worst case, all configurations in the system (finite set) will be executed, but in practice often much fewer executions are needed.
\looseness=-1

Our algorithm will produce the same partitions independent of the order in which configurations are executed.
All subspaces that are derived during any execution of the taint analysis will be derived at some point, because ($1$) we eventually explore all paths in each region and ($2$) we update the partition of each region with the commutative cross-product operation.
\looseness=-1

Theoretically, we can measure performance while running the taint analysis.
In practice though, running a dynamic taint analysis with control-flow tracking imposes significant overhead,
which significantly alters measurement results.
A second reason for separating performance measurement from taint tracking is an optimization to perform taint tracking on a different (smaller) workload than the one used for performance measurement, which we discuss in Sec.~\ref{decisions-sec}.
\looseness=-1

\subsection{Measuring Performance of Regions}
\label{profile-sec}

We measure the execution time of each region when executing a configuration, resulting in performance measurements for each pair of configuration and region.
To this end, we
use an off-the-shelf profiler.
We measure \emph{self-time} per region to track the time spent in the region itself, 
which excludes the time of calls to execute code from other regions. 
\looseness=-1

Since we measure performance in a separate step than the iterative analysis, we can pick a new and possibly smaller set of configurations 
to explore all paths per region.
Ideally, we want to find a minimal set of configurations, such that we have at least one configuration per subspace of each region's partition. 
Since finding the optimal solution is NP-hard\footnote{
The problem can be reduced to the set cover problem, in which the union of a collection of subsets (all subspaces) equals a set of elements called "the universe" (the union of all subspaces).
The goal is to identify the smallest sub-collection whose union equals the universe.
\looseness=-1
},
and existing heuristics from combinatorial interaction testing~\cite{AKTLS:GPCE16, HMGB:IST16, KKL:ICT13, HBG:SRE11} are expensive, we developed our own simple greedy algorithm: Incrementally intersecting subspaces that overlap in at least one configuration, until no further such intersections are possible. 
Then, we simply pick one configuration from each subspace.
However, we never found a smaller set of configurations than the one used in the iterative analysis. 
Thus, we use the same set of configurations to measure performance.
\looseness=-1

\paragraph*{Example}
Four configurations 
cover all subspaces for the three regions of our 
example, 
e.g., $\{\}, \{$\texttt{A}$\}, \{$\texttt{C}$\}, \{$\texttt{A},\texttt{B},\texttt{C}$\}$.
\looseness=-1


\subsection{Building Performance-Influence Models}
\label{build-sec}

In the final step, we build performance-influence models for each region based on ($1$) the partitions identified per region and ($2$) the performance measured per region and observed configuration.
Then, we compose the local models into a performance-influence model for the entire program.
\looseness=-1

Since we collect at least one 
measurement per distinct path through a region, building 
models is straightforward.
For a region with a partition and a set of configurations with corresponding performance measurements, we associate each measurement with the subspace of the partition to which the configuration belongs.
If multiple measured configurations belong to the same subspace of the region's partition, we expect the same performance behavior for that region (modulo measurement noise) and average the measured results.
As a result, we can map each subspace of a region's 
partition to a performance measurement.
For instance, for region \texttt{main} in our example, we report an execution time of $2$ seconds for configurations in $\constr{\texttt{A}}$ and $3$ seconds for configurations in $\constr{\neg \texttt{A}}$.
\looseness=-1

For interpretability, to highlight the influence of options and avoid sets or propositional formulas, we write 
linear models in terms of options and interactions,
for example \verb|m|$_{\texttt{main}} = 3 - 1\texttt{A}$.
\looseness=-1

The global performance-influence model is obtained simply by aggregating all local models; we add the individual influences of each model.
Note that local models can be useful
for understanding and debugging individual regions, as they describe the performance behavior of each region.
\looseness=-1

\begin{figure*}[t]
\begin{center}
  \begin{minipage}[t]{0.5\columnwidth}
    \includegraphics[width=\textwidth]{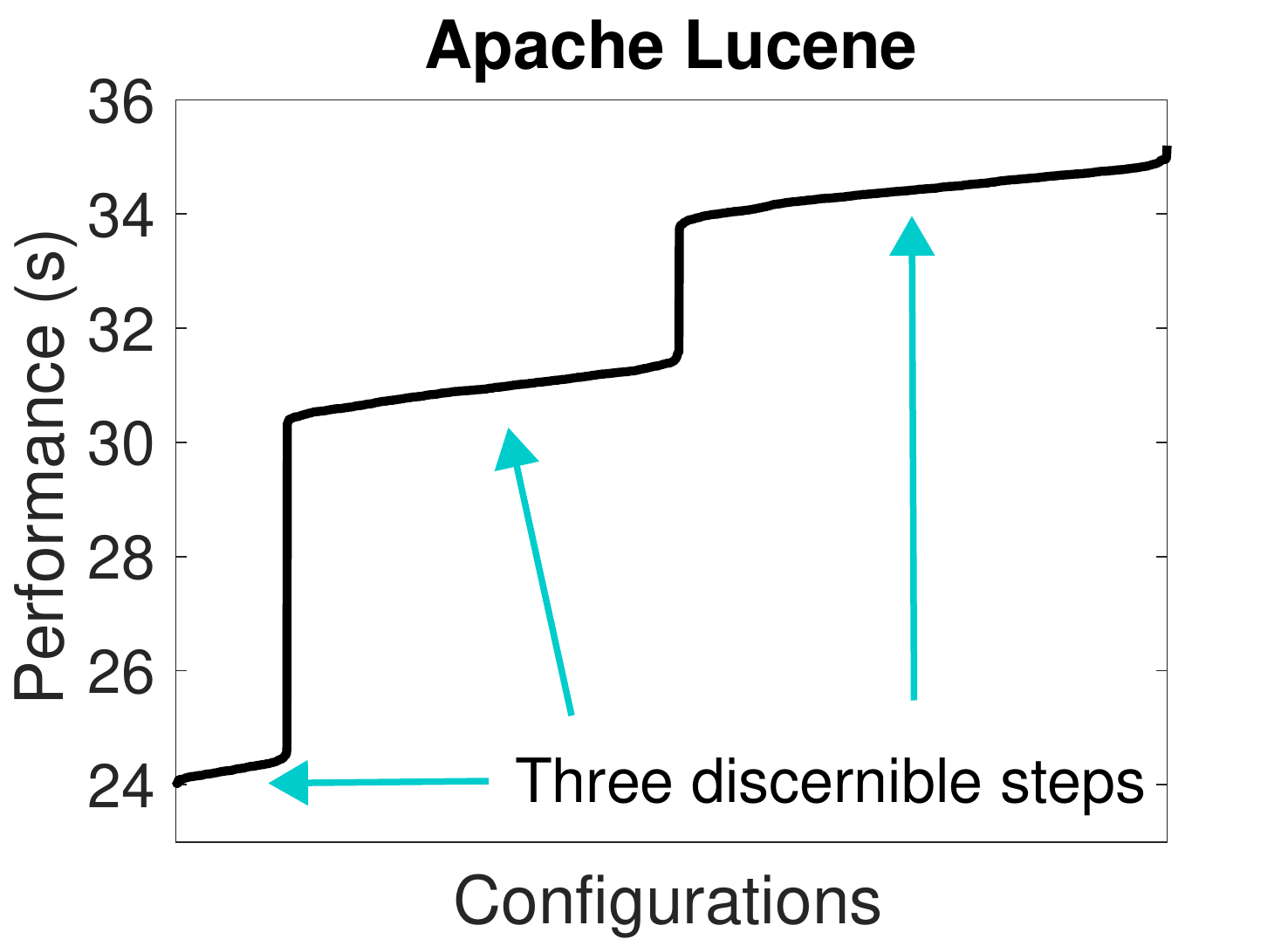}
  \end{minipage}
    \hfill
  \begin{minipage}[t]{0.5\columnwidth}
    \includegraphics[width=\textwidth]{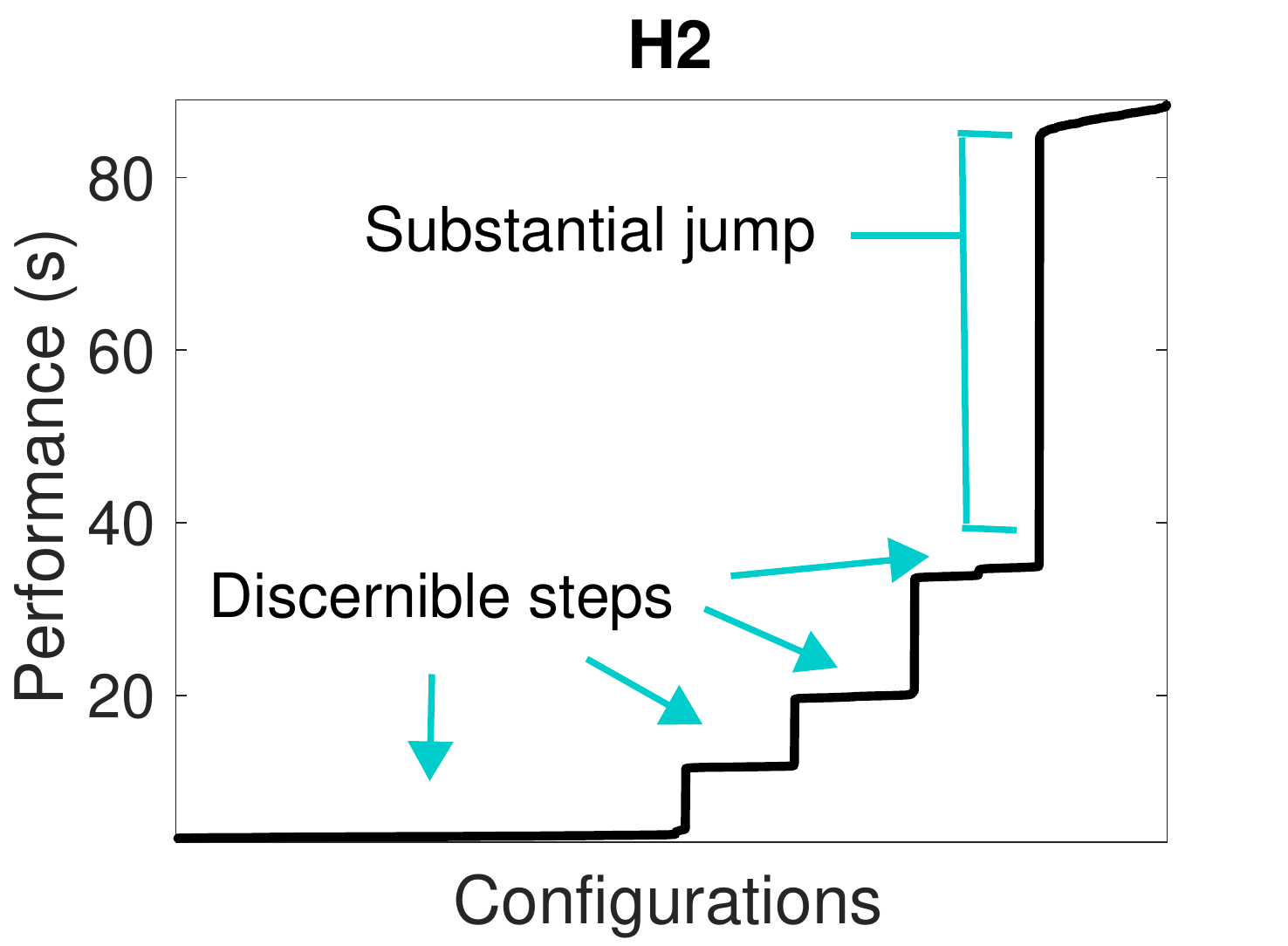}
  \end{minipage}
    \hfill
  \begin{minipage}[t]{0.5\columnwidth}
    \includegraphics[width=\textwidth]{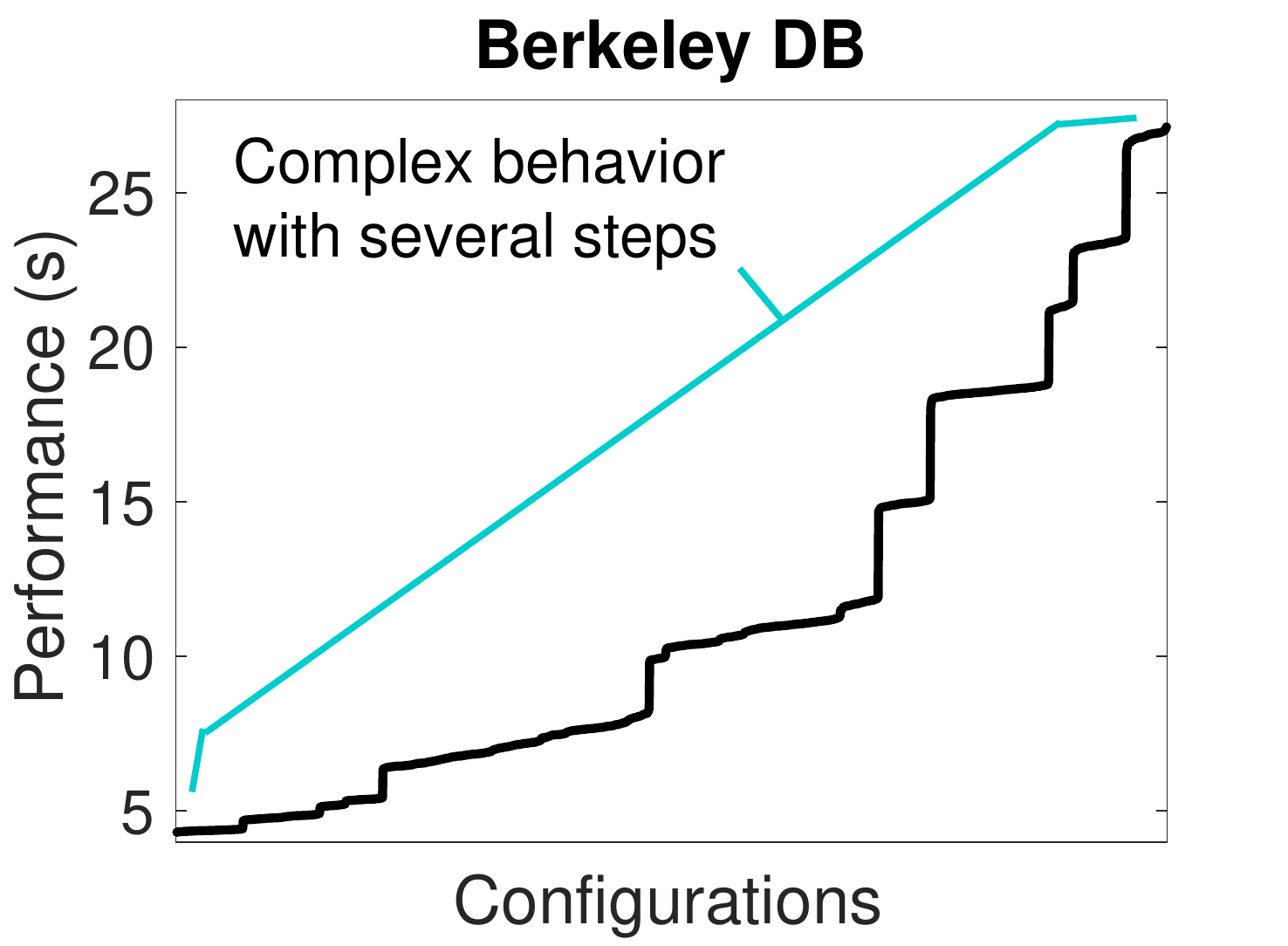}
  \end{minipage}
  \hfill
  \begin{minipage}[t]{0.5\columnwidth}
    \includegraphics[width=\textwidth]{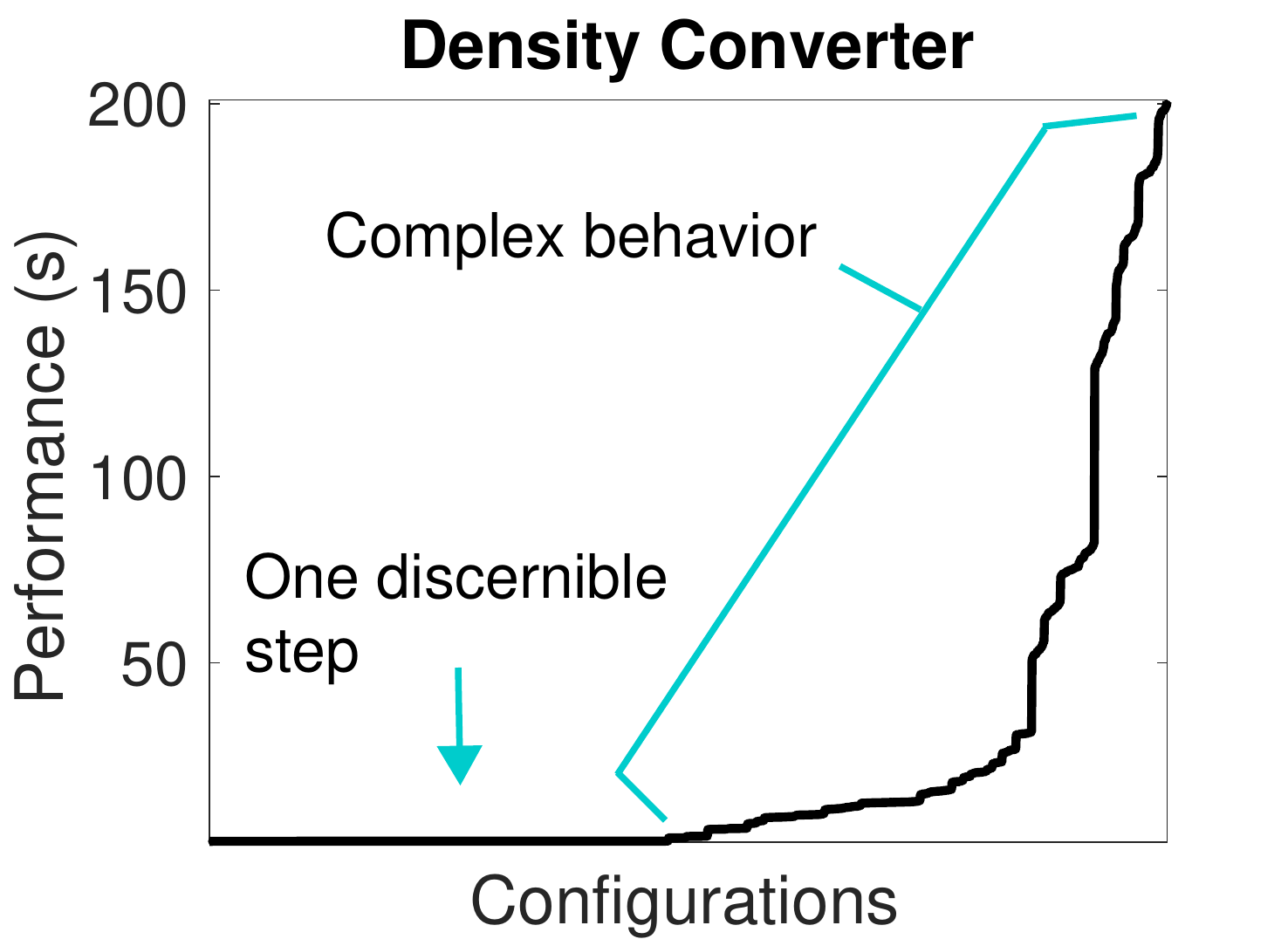}
  \end{minipage}
\end{center}
\caption{Performance behaviors of our subject systems, from simple to complex.
We randomly selected $2000$ configurations and sorted their execution time from fastest to slowest.
\looseness=-1}
\label{perf-beh-fig}
\end{figure*}

\paragraph*{Example}
In our running example, we derive the local 
models \verb|m|$_{\texttt{main}} = 3 - 1\texttt{A}$, \verb|m|$_{\texttt{foo}} = 1\texttt{A} + 3\texttt{AB}$, and \verb|m|$_{\texttt{bar}} = 5 + 15\texttt{A} + 10\texttt{C} + 30\texttt{AC}$, which can be composed
into the global performance-influence model
\verb|m| $=$ \verb|m|$_{\texttt{main}} +$ \verb|m|$_{\texttt{foo}} +$ \verb|m|$_{\texttt{bar}} = 8 + 15\texttt{A} + 10\texttt{C} + 3\texttt{AB} + 30\texttt{AC}$.
\looseness=-1

\section{Design, Optimizations, and Implementation}
\label{decisions-sec}

\paragraph*{Dynamic Taint Analysis Overhead}
\APPROACH\ executes a dynamic taint analysis with control-flow tracking, which imposes significant overhead in the system's execution.
For instance, one execution of our subject system Berkeley DB 
takes about $1$ hour with the taint analysis, whereas about $300$ configurations can be executed in the same time!
In general, we observe $26\times$ to $300\times$ overhead from taint tracking, which varies widely between systems.
To reduce cost, we separate the iterative analysis and performance measurement in two steps and perform the 
former with a \emph{drastically} reduced workload size.
\looseness=-1

This optimization is feasible when the 
workload is \emph{repetitive} and 
repetitions of operations are affected similarly by options, which we conjecture to be common in practice. 
Many performance benchmarks 
execute many 
operations, which are similarly affected by configuration options.
For instance, Berkeley DB's 
benchmark populates a database, where options determine, for example, whether duplicates are allowed and the durability characteristics of a transaction.
The benchmark can be scaled by
a parameter that controls the number of entries to insert, but does not affect which
\emph{operations} are performed.
In our evaluation,
we show that \APPROACH\ generates accurate models using a significantly smaller workload in the iterative analysis.
\looseness=-1

We conduct the actual performance measurements with the original workload to record realistic performance.
\looseness=-1

\paragraph*{Granularity of Regions}
Our approach can be applied to different levels of granularity, but with different tradeoffs.
On one extreme, we could consider the entire system as a single region, but would not benefit from compression.
At the other extreme, we could consider each control-flow statement as the start of its own region, ending with its immediate postdominator,
which allows maximum compression, but results in excessive measurement cost. 
Note the latter is analogous to using an \emph{instrumentation profiler}, but instead of focusing on few locations of interest, as usually recommended~\cite{L:MJP11}, one would add instrumentation throughout the entire system at control-flow statements. 
In our evaluation,
we show that considering each method as a region is a practical compromise.
\looseness=-1

When considering methods as regions, 
we may lose some compression potential compared to more fine-grained regions, 
if 
multiple control-flow statements within a method are influenced by distinct options.
On the other hand, 
we can use off-the-shelf \emph{sampling profilers} that accurately capture performance 
with low overhead, and simply map the performance of methods to the closest regions on the calling stack.
Our empirical evaluation 
shows that method level regions do not require more configurations compared to more fine-grained regions, but significantly reduce the measurement overhead.
\looseness=-1

\paragraph*{Implementation}
We implemented \APPROACH\ for Java systems~\cite{VJSAK:ICSE21SM}.
We used Phosphor, a state of the art tool for dynamic taint analysis~\cite{BK:SIGPLANNOT14} to track options.
We 
annotated the APIs
to load 
options as sources and 
control-flow statements as sinks.
We encoded subspaces as propositional formulas and use SAT$4$J~\cite{SAT4J} for operations on those formulas.
We used JProfiler $10.1$~\cite{JPROFILER10} to measure performance at method level.
\looseness=-1

\section{Evaluation}
\label{eval-sec}

To evaluate the efficiency and effectiveness of our approach, we compare \APPROACH\ to state of the art performance-influence modeling approaches for configurable systems, 
in terms of the cost to generate the models and their accuracy, and discuss their interpretability. 
We evaluate the efficiency of compression
by choosing regions at different granularities.
Specifically, we address the following research questions:
\looseness=-1

\vspace{1ex}

\textit{\textbf{RQ1: How does \BFITAPPROACH\ compare to state of the art performance modeling approaches in terms of cost, accuracy, and interpretability?}}
\looseness=-1

\vspace{1ex}

\textit{\textbf{RQ2: How efficient is compression at different granularities to reduce the cost to build 
models?}}
\looseness=-1

\subsection{Experiment Setup}
\label{experimental-setup-sec}

\paragraph*{Subject Systems}
We selected $4$ configurable, widely-used, open-source Java systems that satisfy the following criteria (common in our domain): (a)~medium- to large-scale systems with over $40$K SLOC and over $20$ options, (b)~systems with binary and non-binary options, (c)~systems with fairly stable execution time (despite nondeterminism and concurrency, we observed execution times within usual measurement noise for repeated execution of the same configuration), and (d)~systems with different performance behaviors (see Fig.~\ref{perf-beh-fig}).
Table~\ref{systems-tbl} provides an overview of all subject systems.
\looseness=-1

We focus on a large subset of all options that are potentially relevant for performance.
We considered options for which the systems' documentation
indicated that they would affect performance, but excluded options that might not influence performance, (e.g., \texttt{-}\texttt{-}\texttt{help}).
This selection is representative of common use cases where users are interested in the performance behavior of many, but not all options.
Following the evaluation of state of the art approaches \cite{SKKABRS:ICSE12, SRKKAS:SQJ12, SGAK:ESECFSE15, KGSGA:ICSE19, SGSAC:ASE15, GYSASVCWY:ESE17, LKB:TSE18, SRA:GPCE13,  MKRGA:ICSE16, HNADPB:ESE18}, we selected, for non-binary options, two different values and encoded the values as a binary option.
\looseness=-1

\begin{table}[t]
\scriptsize
\centering
\caption{Subject systems.}
\begin{threeparttable}
\begin{tabular}{llrrrr}
\toprule
System & Domain & \#KLOC & \#Opt. & \#Conf. & V/ID \\ \midrule
Apache Lucene & Index/Search & 396 & 17 & 131K & \texttt{7.4.0} \\
H2 & Database & 142 & 16 & 65K & \texttt{1.4.201} \\
Berkeley DB & Database & 164 & 16 & 65K & \texttt{7.5.11} \\
Density Converter & Image processor & 49\tnote{1} & 22 & 4.9M & \texttt{110c4} \\
\bottomrule
\end{tabular}
Opt: Options;
Conf: Configurations;
V/ID: Version/Commit ID; $^1$: Interface to several libraries for processing images with $1.5$K SLOC, included in the evaluation, as the system has a large configuration space.
\looseness=-1
\end{threeparttable}%
\label{systems-tbl}
\end{table}

We executed a long-running benchmark shipped with the system, representative of a user analyzing the system's performance under different configurations.
Detailed information about the 
benchmarks 
is included in our appendix~\cite{VJSAK:ICSE21SM}.
\looseness=-1

For each system, we changed the workload parameter to run the iterative dynamic taint analysis with a smaller workload by factors ranging from $20$ to $50000$, depending on the system.
The smaller workload reduced the iterative analysis time for a single configuration, for example, on average, from $1$ hour to $5$ seconds for Berkeley DB.
Information about the 
changes in the workload can be found in our appendix~\cite{VJSAK:ICSE21SM}.
\looseness=-1

\paragraph*{Hardware} 
Analyses and measurements were executed on an Ubuntu $18.04$ LTS desktop, with a $3.4$ GHz $8$-core Intel Core i$7$ processor, $16$ GB of RAM, and Java HotSpot\textsuperscript{TM} $64$-bit Server VM (v$1.8.0$\_$202$).
\looseness=-1


\paragraph*{Performance Measurement} 
We established a
large evaluation set for quantifying accuracy by measuring the performance of $2000$ randomly selected configurations, as the configuration spaces of the subject systems are intractably large.
We executed each configuration five times and used the median
to reduce the effects of measurement noise.
We initiated one VM invocation per configuration, thus all measures include startup time~\cite{GBE:SIGPLANNOT07}.
For \APPROACH, we profiled each system with JProfiler's default sampling rate of $5$ms.
\looseness=-1

\subsection{RQ1: Comparison to the State of the Art}
\label{rq1-comparison-sec}

With RQ1, we evaluate the cost to generate performance-influence models with \APPROACH, at method granularity, asses their accuracy and interpretability, and compare them to state of the art approaches. 
Specifically, we compare \APPROACH\ to 
numerous combinations of sampling and learning approaches.
For learners, we evaluate variations of linear regressions~\cite{SKKABRS:ICSE12, SRKKAS:SQJ12, SGAK:ESECFSE15}, decision trees and random forests~\cite{GCASW:ASE13, SGSAC:ASE15, GYSASVCWY:ESE17, GSA:PPSCNSB19}, and a neural network.
For sampling, we evaluate uniform random sampling with $50$ and $200$ configurations, feature-wise sampling (i.e., enable one option at a time), and pair-wise sampling (i.e., cover 
all combinations of all pairs of options)~\cite{MKRGA:ICSE16}.
We selected $50$ and $200$ random configurations to use more configurations than other sampling strategies and use sampling sets comparable
to 
ones used in related research.
Due to space restrictions, we report results only for stepwise linear regression and random forest.
The other results can be found in our appendix~\cite{VJSAK:ICSE21SM}. 
\looseness=-1

Note that we do not compare against approaches for selecting the fastest configuration~\cite{KGSGA:ICSE19, ZLGBMLSY:SOCC17, OBMS:ESECFSE17}, as those approaches solve a pure optimization problem where modeling the entire configuration space is not necessary.
\looseness=-1

We do not evaluate existing white-box approaches due to their limitations~\cite{SRA:GPCE13, VJSSAK:ASEJ20}.
None of the approaches could analyze any of our subject systems, except for a subset of Density Converter's code base, which we discuss in Sec.~\ref{related-work-sec}.
\looseness=-1

\begin{table}[t]
\centering
\caption{Cost comparison.}
\begin{subtable}[t]{\columnwidth}
\centering
\scriptsize
\begin{threeparttable}
\begin{tabular}{lrrrr}
\toprule
Sample & Apache Lucene  & H2 & Berkeley DB & Density Conv. \\ \midrule
BF & 2\textsuperscript{17} [\textasciitilde48.4d] & 2\textsuperscript{16} [\textasciitilde16.0d] & 2\textsuperscript{16} [\textasciitilde8.6d] & 2\textsuperscript{22} [\textasciitilde3.6y] \\
R$50$  & 50 [26.7m] & 50 [16.4m] & 50 [9.1m] & 50 [10.1m] \\
R$200$ & 200 [1.8h] & 200 [1.1h] & 200 [36.4m] & 200 [40.4m] \\
FW & 17 [8.6m] & 16 [2.4m] & 16 [4.8m] & 22 [8.2m] \\
PW & 154 [1.3h] & 137 [21.1m] & 137 [39.4m] & 254 [1.8h] \\
\APPROACH & 26 [14.9m] & 64 [22.6m] & 144 [30.2m] & 88 [16.6m] \\
\bottomrule
\end{tabular}
The time to measure configurations for brute-force (BF)
is extrapolated from $2000$ randomly selected configurations.
\looseness=-1
\caption{Cost of sampling configurations.}
\label{rq1-sample-cost-tbl}


\looseness=-1
\end{threeparttable}%
\end{subtable}

\bigskip

\begin{subtable}[t]{\columnwidth}
\centering
\scriptsize
\begin{threeparttable}
\begin{tabular}{lrrrr}
\toprule
Approach & Apache Lucene & H2 & Berkeley DB & Density Conv. \\ \midrule
R$50$ \& LR & 8.9s & 6.6s & 5.7s & 14.9s \\
R$200$ \& LR & 6.8m & 4.6m & 4.9m & 1.6m \\
FW \& LR & 9.4s & 4.3s & 7.7s & 19.8s \\
PW \& LR & 1.7m & 44.8s & 3.6m & 5.5m  \\
* \& RF & $\le$0.2s & $\le$0.2s & $\le$ 0.3s & $\le$0.2s \\
\APPROACH & 28.9m & 9.3m & 11.2m & 8.5m \\
\bottomrule
\end{tabular}
\caption{Learning/Analysis time.}
\label{rq1-analysis-cost-tbl}

\medskip

BF: Brute Force;
R$50$: $50$ random configurations;
R$200$: $200$ random configurations;
FW: Feature-wise;
PW: Pair-wise;
LR: Stepwise linear regression;
RF: Random forest;
*: Results stable across all samples.
\looseness=-1
\end{threeparttable}%
\end{subtable}
\label{rq1-cost-tbl}
\end{table}

\paragraph*{Cost Metric}
We report the number of configurations executed 
to generate a model and time to measure
configurations.
For the learning approaches, we 
report the learning time.
For \APPROACH, we 
report
the 
time to execute the iterative analysis.
\looseness=-1

\paragraph*{Accuracy Metric} 
We report the Mean Absolute Percentage Error (MAPE), which measures the mean difference between the values predicted by a model and the values actually observed (i.e., the baseline), lower is better.
\looseness=-1

As profiling the performance of regions with our approach adds some overhead 
(about $8\%$ in our subject systems), we expect 
our models 
to be systematically biased. 
We performed an optional linear correction to account for the average overhead in our subject systems, learned from measurements with and without profiling of $5$ randomly selected configurations. 
We separately report these "corrected" MAPE values in our results.
\looseness=-1


\begin{table}
\scriptsize
\centering
\caption{\footnotesize MAPE comparison (lower is better).}
\begin{threeparttable}
\begin{tabular}{lrrrr}
\toprule
Approach & Apache Lucene & H2 & Berkeley DB & Density Conv. \\ \midrule
R$50$ \& LR & \bl \textbf{4.5} & 124.1 & 19.7 & 1037.2  \\
R$200$ \& LR & \bl \textbf{2.9} & 93.9 & 14.9 & 434.5 \\
FW \& LR & \bl \textbf{7.9} & 129.3 & 768.7 & 1596.0 \\
PW \& LR & \bl \textbf{4.7} & 113.3 & 34.2 & 1596.0 \\
R$50$ \& RF & \bl \textbf{0.4} & \bl \textbf{1.1} & 16.4 & 59.9 \\
R$200$ \& RF & \bl \textbf{0.3} & \bl \textbf{0.7} & \bl \textbf{1.1} & \bl \textbf{5.5} \\
FW \& RF & \bl \textbf{8.7} & 119.0 & 106.1 & 1185.9 \\
PW \& RF & \bl \textbf{4.0} & 124.6 & 46.9 & 27.3 \\ 
\APPROACH\ (O) & \bl \textbf{8.0} & \bl \textbf{9.3} & \bl \textbf{6.7} & \bl \textbf{11.2} \\
\APPROACH\ (C) & \bl \textbf{3.2$\pm$0.2} & \bl \textbf{2.9$\pm$0.5} & \bl \textbf{5.0$\pm$0.6} & \bl \textbf{9.4$\pm$2.4\tnote{1}} \\
\bottomrule
\end{tabular}
LR: Stepwise linear regression;
RF: Random forest;
R$50$: $50$ random configurations;
R$200$: $200$ random configurations;
FW: Feature-wise;
PW: Pair-wise;
O: Original model;
C: Model corrected for profiler overhead, reporting mean and standard deviations over 5 corrections.
\textbf{Bolded} values in \colorbox{\colorbl}{cells} indicate similarly low errors.
\looseness=-1
\begin{tablenotes}
\item[1] Non-linear profiling overhead might have increased the error~\cite{WAS:ICSE21}.
\end{tablenotes}
\end{threeparttable}%
\label{rq1-mape-tbl}
\end{table}

\paragraph*{Interpretability} 

We intend our models to also be used in understanding and debugging tasks (see Sec.~\ref{goals-subsec}).
Unfortunately, measuring interpretability of models is nontrivial and controversial.
In the machine learning community, interpretability is an open research problem with an active community, but without a generally agreed measure or even definition for interpretability~\cite{M:IML20,DK:ARCHIVE17}.
\looseness=-1

Generally, interpretability captures the ability of humans to make predictions, understand predictions, or understand the decisions of a model~\cite{M:IML20}.
When the model is complex (e.g., the model includes numerous decisions), humans have more difficulty understanding the model directly.
Some simpler forms of models are usually considered \emph{inherently interpretable}, because humans can inspect and understand the models directly.
For example, scientists have decades of experience using and interpreting linear models (e.g., much of empirical software engineering research relies on interpreting linear model coefficients).
Models with more complex structures and very large numbers of decisions (e.g., deep neural networks with
millions of weights or random forests with hundreds of trees), exceed human capacity for directly understanding the model.
With these more complex models, the trend is to develop \emph{post-hoc explanations}, where tools provide explanations for specific aspects of the model (e.g., the reason for a given prediction) without having to understand the model's internals~\cite{M:IML20,RSG:KDD16,LL:ANIPS17,SK:KIS14}. 
The use of post-hoc explanations is, however, controversial, as the explanations usually are
only approximations that may be unreliable or even misleading~\cite{R:NMI19}.
\looseness=-1

We do not attempt to quantify interpretability.
Instead, we generally consider sparse linear models as inherently interpretable (assuming a moderate numbers of terms).
This is supported by prior interviews that have shown that developers understand linear performance-influence models with a few dozen terms~\cite{KSKGA:SOSYM18}. 
By contrast, we consider random forests and neural networks as not inherently interpretable, and do not evaluate the reliability or usefulness of post-hoc explanations.

To assure readers that the linear models that we produce are indeed \emph{sparse} and, hence, likely interpretable by humans, we report the number of terms they contain.
As our algorithm to build these models (Sec.~\ref{build-sec}) does not include any machine learning and regularization, the algorithm detects and reports even minuscule amounts of measurement noise.
Hence, we exclude all trivial terms that do not make meaningful contributions to the systems' performance.
We report the number of terms (options or interactions) that contribute, at least, $0.3$ seconds, which is approximately $1\%$ of the execution time of the default configurations of our subject systems.
Our appendix contains additional data at other thresholds~\cite{VJSAK:ICSE21SM}.



\paragraph*{Results} 
We report the main results in Table~\ref{rq1-cost-tbl} (cost) and~\ref{rq1-mape-tbl} (accuracy), and our
appendix~\cite{VJSAK:ICSE21SM} includes results for additional learners (neural networks, decision trees).
Overall, \APPROACH\ builds models that are similarly accurate to those learned by the most accurate and expensive black-box approach (random forests with 200 samples), but our models are interpretable and usually built more efficiently, despite the cost of the additional taint analysis step.
\APPROACH\ outperforms other approaches that build linear models by a wide margin.

While random forest with $200$ samples produced slightly more accurate models than \APPROACH, our approach was usually 
more efficient, in some cases building models in half the time, while also generating local and interpretable models (see Fig.~\ref{mape-cost-interpret-h2-fig}).
The efficiency originates from \APPROACH's white-box analysis to identify a small number of relevant configurations to capture the performance-relevant interactions.
By contrast, as our results show, black-box approaches perform significantly worse on such small samples (e.g., compare R50 and R200 results).
\looseness=-1

The linear models produced by \APPROACH\ are moderate in size, with $19$, $16$, $32$, and $72$ performance-relevant terms (options or interactions) for Lucene, H2, Berkeley DB, and Density Converter respectively.
Our models are similar in size to models learned with linear regression from samples (e.g., $26$, $13$, $28$, and $30$ terms using R$200$ \& LR), but much more accurate.
At this size, we argue that manual inspection of the models is still plausible.
More importantly, the performance influences can be mapped to $24$, $5$, $15$, and $9$ specific regions in the code for the four systems respectively.
\looseness=-1

\begin{framed}
	\noindent{
	\textit{RQ1: 
	In summary, models produced with \ITAPPROACH\ have comparable
	accuracy to the most accurate and expensive black-box approaches, but can often be built more efficiently.
	Additionally, the models significantly outperform other black-box approaches that produce (interpretable) linear models. 
	The models are interpretable and can be mapped to specific code regions.
	\looseness=-1
	}
	}
\end{framed}

\subsection{RQ2: Compression Efficiency}
\label{rq2-compress-sec}

With RQ2, we explore the impact of choosing regions at different granularities on the efficiency of \APPROACH, both in terms of the \emph{number of configurations} to measure and the \emph{overhead} to perform these measurements.
\looseness=-1

\paragraph*{Procedure} 
For RQ1, we executed the taint analysis
considering each 
method
as a region.
We additionally tracked 
partitions 
for 
control-flow statements
and derived partitions for the entire program
by
combining the partitions of all methods.
\looseness=-1

\paragraph*{Result -- Number of Configurations}
In Table~\ref{rq2-compression-tbl}, we report the size of the minimum set of configurations needed to cover each subspace of each region's partition
for each granularity.
When considering the entire program as a region, significantly more configurations need to be explored, as we do not benefit from compression.
Interestingly though, while there are, as expected, fewer regions at the method level than at the control-flow statement level, the number of configurations needed is the same.
These results show that compression at finer-grained levels than the method level do not yield additional benefits in our subject systems.
\looseness=-1

We found that the control-flow statement regions combined within a method are usually 
partitioned in the same way.
Only in $3$ out of the $2263$ method-level regions, 
the method's partition had more subspaces than the corresponding control-flow statement regions
(e.g., 
two \texttt{if} statements depending on different options).
However, in all three cases, the additional subspaces were already explored in other parts of the program.
Hence, no additional configurations needed to be explored.
\looseness=-1

We conclude
that 
fined-grained compression 
is highly effective, but that control-flow granularity does not seem to offer significant compression benefits over method granularity.
\looseness=-1

\emph{Results -- Measurement Overhead:}
Measuring performance at different granularities requires different strategies, each with vastly different amounts of measurement overhead.
Measuring at the program level is cheap, as a single end-to-end measurement is sufficient to measure the entire program (e.g., Unix \texttt{time}). 
At finer granularities, multiple measurements of different parts of the program are required.
Instrumenting the program at control-flow statements and corresponding post-dominators leads to significant measurement overhead, as the measurement instructions are executed 
frequently
(similar to an instrumentation profiler).
For instance, in the time to measure methods
(e.g., Berkeley DB executed $144$ configurations in $30.2$ minutes), not a single configuration finished measuring control-flow statements.
\looseness=-1

By contrast, measuring the performance of numerous methods is inexpensive with a \emph{sampling profiler} (cf. \APPROACH\ in Table~\ref{rq1-sample-cost-tbl}), for which we observed a mostly linear overhead of about $8\%$ in our subject systems.
\looseness=-1

\begin{table}[t]
\scriptsize
\centering
\caption{Number of regions and configurations to measure with compression at different region granularities.}
\begin{threeparttable}
\begin{tabular}{lrrrrrr}
\toprule
 & \multicolumn{2}{c}{Control-flow} & \multicolumn{2}{c}{Method} & \multicolumn{2}{c}{Program} \\
 \cmidrule(rl){2-3}
 \cmidrule(rl){4-5}
 \cmidrule(rl){6-7}
System & \#Reg. & \#Conf. & \#Reg. & \#Conf. & \#Reg. & \#Conf. \\ \midrule
Lucene & 1654 & \bl \textbf{26} & 551 & \bl \textbf{26} & 1 & 16384 \\
H2 & 2483 & \bl \textbf{64} & 932 & \bl \textbf{64} & 1 & 256 \\
Berkeley DB & 2152 & \bl \textbf{144} & 718 & \bl \textbf{144} & 1 & 2048 \\
Density Converter & 190 & \bl \textbf{88} & 62 & \bl \textbf{88} & 1 & 4608 \\
\bottomrule
\end{tabular}
\#Reg: Number of regions;
\#Conf: Number of configurations.
\textbf{Bolded} values in \colorbox{\colorbl}{cells} indicate the minimum number of configurations to cover all partitions' subspaces. 
\looseness=-1
\end{threeparttable}%
\label{rq2-compression-tbl}
\end{table}

\begin{framed}
	\noindent{
	\textit{RQ2: In summary, compression at method and control-flow granularities is highly efficient to reduce measurement effort.
Compression at method granularity provides a good compromise between compression potential and measurement overhead.
	\looseness=-1
	}
	}
\end{framed}


\subsection{Limitations and Threats to Validity}
\label{threats-sec}

\paragraph*{Limitations of the Taint Analysis}
\label{rq2-small-vs-large-idta-sec}
Our approach to compute partitions per region may produce inaccurate results due to two sources of inaccuracy:
($1$) A standard dynamic taint analysis cannot reason about \emph{paths not taken}~\cite{AF:PLAS09} and, thus, may miss some taints, and ($2$) our
use of a small workload to reduce the cost of 
the iterative analysis may lead to some missed interactions.
The former threat is somewhat mitigated by 
exploring multiple configurations and using the cross product when updating partitions; 
we 
see both branches of each control-flow decision.
The latter issue depends on how the workload is shortened (see Sec.~\ref{decisions-sec}), but will likely have a low impact in highly repetitive workloads.
Both threats can 
lead to generating inaccurate models.
\looseness=-1

Importantly, our results for RQ$1$ suggest that inaccuracies on the regions' partition caused by these threats resulted in, at most, minor accuracy degradation in our performance-influence models, given the consistently high accuracy achieved across all subject systems.
\looseness=-1

Furthermore, we used a debugging strategy to identify potential effects of inaccurate partitions.
As discussed in Sec.~\ref{build-sec}, 
multiple executions of configurations within the same subspace of a region's partition must have the same performance behavior.
Significant 
differences,
beyond normal measurement noise, indicate that the region's partition might be inaccurate and not capturing all relevant interactions.
\looseness=-1

Specifically, we analyzed the performance measurements of all regions, searching for regions with a significant performance influence ($>0.1$ms in any observed configuration) and with a high variance among the execution times of the same subspace in a region (coefficient of variation of the execution times $>1.0$).
Among the $2263$ method-level regions that we analyzed, 
we found only $8$ of such regions, all in Lucene.
\looseness=-1

We executed the iterative analysis with the regular workload, which resulted in additional subpaces in $7$ out of $8$ regions, due to slightly different taints in the shorter workload.
While the missing subspaces
slightly decreased the MAPE from $8.0$ to $7.4$ (as a result of slight changes in the coefficients in the model, not from new options nor interactions becoming performance relevant), 
the time to run the taint analysis with the regular workload is \emph{extremely} expensive; $11$ hours instead of $29$ minutes to run the same $26$ configurations.
In fact, the iterative analysis did no finish executing after $24$ hours in the other subject systems!

We argue that the extremely high cost and inability to use the regular workload 
does not outweigh a potential slight accuracy increase 
of the already highly accurate models that we generated.
These results support our conjecture that inaccuracies on the regions' partition caused by using an unsound analysis with a small workload are only a minor issue in practice.
\looseness=-1

\paragraph*{Threats to Validity}
Beyond the 
limitations of the taint analysis,
measurement noise cannot be excluded and may affect 
\emph{all} results.
We reduced this threat by repeating measurements 
on a dedicated machine and using the median.
\looseness=-1

The primary threat to external validity is the selection of subject systems.
While we selected medium- to large-scale widely-used open-source Java configurable systems from different domains, readers should be careful when generalizing results. 
For instance, all analyzed systems are multi-threaded, but had mostly deterministic performance behavior.
Additionally, we analyzed
a single 
configurable system, whereas systems composed of numerous configurable systems, deployed in distributed environments, and implemented in different languages are beyond the scope of this paper.
\looseness=-1

Another threat is the selected subset of options,
which might not affect performance at all, making modeling a trivial task.
We selected options for which the systems' documentation or the options’ functionality indicated that they would affect performance, and we observed a wide range of execution times for the configurations that we measured (see Fig.~\ref{perf-beh-fig}).
\looseness=-1

\section{Related Work}
\label{related-work-sec}

In Sec.~\ref{state-of-the-art-sec}, we discussed performance modeling in general and evaluated closely related state of the art black-box approaches for performance-influence modeling to \APPROACH.
In this section, we discuss \emph{additional} research to position \APPROACH\ in a broader context of prior work.
\looseness=-1

Only few researchers have explored white-box performance modeling of configurable systems in prior work~\cite{SRA:GPCE13,VJSSAK:ASEJ20,LWHL:EUROSYS20}.
\citet{SRA:GPCE13} introduced the idea of white-box analysis, but assumed a specific programming style that provided a static mapping from options to regions and ignored data-flow between regions, severely limiting the systems that could be analyzed.
In prior work, we developed ConfigCrusher~\cite{VJSSAK:ASEJ20}, which used static data-flow analysis to map options to regions.
ConfigCrusher also measures the performance of regions and builds local performance-influence models.
ConfigCrusher is, however, limited by the overhead of the static analysis and was unable to scale beyond small systems (e.g., it never terminated when we tried it on our subject systems).
By contrast, \APPROACH\ uses a dynamic data-flow analysis that scales to large systems, demonstrating the feasibility of white-box approaches to efficiently build accurate performance-influence models for large-scale configurable systems.
For comparison, we evaluated both approaches on Density Converter, though \APPROACH\ analyzed all used libraries, resulting in comparable accuracies ($4.3$ with ConfigCrusher vs. $9.4$ with \APPROACH\ in terms of MAPE), but measuring fewer configurations ($256$ with ConfigCrusher vs. $88$ with \APPROACH).
\looseness=-1

In prior work~\cite{VJSSAK:ASEJ20}, we also adapted \emph{SPLat}~\cite{KMKBSBD:ESECFSE13} for performance modeling, originally designed for \emph{testing} configurable systems.
SPLat uses lightweight instrumentation to observe which options are accessed during execution, and in which order, to explore all combinations of options that it encounters during execution.
This type of analysis is cheaper than our taint tracking, but explores many spurious interactions, resulting in SPLat essentially exploring all configurations.
\looseness=-1

\citet{LWHL:EUROSYS20} used intraprocedural control-flow analysis to identify usage patterns of individual options in the source code for predicting performance properties based on the patterns.
In contrast to our work, they do not measure performance, 
build performance-influence models, 
nor 
consider interactions.
\looseness=-1

While our line of work focuses on the influence of configuration options, which are easy to change by users without modifying the implementation, there is a active research field to identify bottlenecks and performance bugs, generally using different forms of static or dynamic analysis~\cite[e.g.,][]{GFX:ICSE2012,NSML:ICSE13,YP:EMSE18,CLBKMG:ICSECP18,JAH:OOPSLA11,HDGZX:ICSE12,CAPPJ:TACO15,LPSS:ISSTA18}.
For example, \citet{CAPPJ:TACO15} use both techniques to identify performance bottlenecks that can be analyzed and optimized in isolation.
These kinds of approaches are focusing on different concerns than performance-influence models but are likely complementary for developers seeking to understand a system's performance behavior.
\looseness=-1


At the same time, researchers have used static and dynamic analyses to characterize and track options in configurable systems~\cite{NKCFP:FSE16, MWKTS:ASE16, DALC:ISSRE16, SD:JSS18, LKB:TSE18, TD:ECOOP18, XJHZLJP:OSDI16, TAKSS:CSUR14,TD:ECOOP16}.
For example, \citet{TD:ECOOP16} use dynamic taint analysis
to identify the use of stale configuration data.
Our work is inspired by insights from analyzing configurable systems and uses similar analysis strategies as foundations to map options to regions.
\looseness=-1


\looseness=-1

\section{Conclusion}
We presented \APPROACH, a white-box approach that efficiently builds accurate and interpretable performance-influence models for configurable systems, without relying on traditional sampling and learning techniques.
\APPROACH\ employs an iterative dynamic taint analysis to identify where options influence the system, building and composing local linear performance-influence models.
Our empirical evaluation on $4$ systems demonstrates the accuracy and efficiency of \APPROACH.
\looseness=-1

\section{Acknowledgements}
We specially want to thank Katherine Hough and Jonathan Bell for their indispensable help with Phosphor.
We thank Chu-Pan Wong, Jens Meinicke, and Florian Sattler for their comments during the development of this work, the FOSD 2019 meeting participants for their feedback on the iterative dynamic taint analysis, and Claire Le Goues and Rohan Padhye for their feedback on earlier drafts of this paper.
This work has been supported in part by the Software Engineering Institute, NSF (Awards 1552944, 1717022, and 2007202), NASA (RASPERRY-SI 80NSSC20K1720), the German Federal Ministry of Education and Research (BMBF, 01IS19059A and 01IS18026B), and the German Research Foundation (DFG, SI 2171/2, SI 2171/3-1, AP 206/11).
\looseness=-1

\footnotesize
\bibliographystyle{myabbrvnat}
\bibliography{bibliography.bib}

\end{document}